# Dynamics of a diathermal versus an adiabatic piston in an ideal gas: Langevin's and phase-space approaches

Bruno Crosignani,[1,2,4] Paolo Di Porto[2,4] and Claudio Conti[3,4]

1. California Institute of Technology, Pasadena, Ca 91125

2. Universita' dell'Aquila, 67010 L'Aquila, Italy

3. Research Center "Enrico Fermi," Via Panisperna 89/A, I-00184, Rome, Italy

4. Research Center Soft INFM-CNR c/o Università di Roma "La Sapienza," I-00185, Rome, Italy

**Abstract**

We present a comparison between the random motion of an adiabatic and a diathermal piston sliding in a perfect gas. In particular, their dynamical behaviour, if investigated by means of Langevin's approach, shows the amplitude of the adiabatic-piston random displacements around its equilibrium position to be much larger (by a factor up to $(M/m)^{1/2}$, where M and m are the piston mass and the mass of the single gas molecule) than that of the diathermal piston. The origin of this intriguing difference, which is accounted for in the frame of Langevin's approach, is also explored in terms of a space-phase analysis.

## 1. Introduction

The adiabatic-piston problem, first pointed out by H. B. Callen in 1960, [1] has been the object of a renewed growing interest in the last ten years. The original problem, which consists in finding the mechanical equilibrium position of a frictionless adiabatic piston sliding in an isolated cylinder of length L, parting it into two sections containing the same number N of molecules of a given perfect gas (see Fig.1), has been first solved in 1996 by using a simple kinetic approach.[2] Successively, after the degree of sophistication of the problem started to be fully appreciated, a number of papers have been published, disclosing the existence of two stages of approach to equilibrium: a first deterministic one, which leads the system in a state of mechanical equilibrium, followed by a stochastic one, driven by microscopic fluctuations, which eventually takes, on a much longer time scale, the system to a state of both mechanical and thermal equilibrium.[3] The second stage has been investigated by means of a number of different techniques and under various assumptions (for an extensive bibliography, see [4]).

The adiabatic-piston evolution process, when starting from an initial condition of both mechanical and thermal equlibrium, has been studied in Ref. [5] by relying on Langevin's approach. It has been shown that in the overdamped regime, corresponding to $mN/M \gg 1$ and under specific hypotheses on both system dimension L and number of molecules N, a stationary equilibrium, i.e., a time-independent distribution function of the piston position, is eventually attained on a time scale $t_{as} = (L/v)M/m$, where v is the most probable velocity of the gas. This situation, although stationary, does not correspond to ordinary thermodynamic equilibrium, since it is associated with position fluctuations larger by a factor $(M/m)^{1/2}$ than the standard thermal ones. They induce, in turn, sizeable random temperature variations in the two sides of the piston. This contradicts the usual statement that any isolated system, left to itself, eventually reaches a final thermodynamic-equilibrium position. One expects the above conclusion to be met with some scepticism; this in turn may question the validity of using Langevin's approach, on which this conclusion is based, to investigate the dynamical evolution of the adiabatic piston.

In this paper, in order to validate our result, we have applied the Langevin approach to study the evolution of both an adiabatic piston and a highly conducting one (diathermal), showing how the results obtained in the latter case are those expected in the frame of ordinary thermodynamic equilibrium. The two pistons differ only by their thermal conductivity, and the successful application of Langevin's method to the diathermal piston (an interesting problem in its own right) suggests its validity also in connection with the adiabatic one. In fact, the main hypothesis justifying its application is the linearity of the underlying equation of motion,[6] which is in turn justified by the smallness of piston position fluctuations as compared to the system dimension L. In order to shed further light on the problem, we have recovered the results obtained for the diathermal piston by adopting a more fundamental approach based on a phase-space analysis associated with the usual ergodicity assumption, i.e., probability density proportional to the phase-space volume element. In this way, we recover the result obtained by means of Langevin's approach, which corroborates its validity. We have as well obtained the same result by maximizing Shannon's entropy for an ensemble of identical diathermal pistons.

## 2. Deterministic stage of piston evolution

In order to apply Langevin's approach to the piston dynamics, it is expedient to identify the deterministic equation it obeys in the presence of the surrounding gas, neglecting fluctuations. Successively, once the piston has reached a mechanical equilibrium position, one has to add the appropriate stochastic Langevin force associated with microscopic fluctuations. Thus, the first step is to identify the appropriate deterministic equation. This problem has been solved by means of a kinetic model in the particular situation of an ideally insulating (adiabatic) piston.[2] The general case, accounting for the presence of a finite piston heat-conductivity κ, has been dealt with by directly using first and second law.[7] As expected, if κ=0, the results of Refs. [2] and [7] coincide (but for the presence of an extra-term proportional to the square of the piston velocity, which can be shown to be negligible in the deterministic regime). More precisely, the equation of motion for the piston position X reads [2]

$$\ddot{X} = \frac{nR}{M}\left(\frac{T_A}{X} - \frac{T_B}{L-X}\right) - \sqrt{\frac{8nRM_g}{\pi M^2}}\left(\frac{\sqrt{T_A}}{X} + \frac{\sqrt{T_B}}{L-X}\right)\dot{X} + \frac{M_g}{M}\left(\frac{1}{X} - \frac{1}{L-X}\right)\dot{X}^2 \,, \quad (1)$$

where $T_A(t)$ and $T_B(t)$ are the temperatures in sections A and B, M the piston mass, n and $M_g$ the common mole number and gas mass in A and B, and R labels the gas constant. Equation (1) is coupled to two evolution equations for $T_A(t)$ and $T_B(t)$ (see Refs. [2] and [7]), explicitly containing the thermal conductivity κ as a parameter, the two extreme cases, diathermal and adiabatic, corresponding to $\kappa \to \infty$ and $\kappa \to 0$, respectively. Hereafter, we assume the piston, initially clamped in a position X(t=0), be free to move starting from t=0.

**Diathermal piston**

The coupled-equation system for X(t), $T_A(t)$ and $T_B(t)$ is consistent, as intuitive, with $T_A(t) = T_B(t) = T(t)$ and, as a consequence, Eq.(1) reduces to

$$\ddot{X} = \frac{nRT}{M}\left(\frac{1}{X} - \frac{1}{L-X}\right) - \sqrt{\frac{8nRM_g T}{\pi M^2}}\left(\frac{1}{X} + \frac{1}{L-X}\right)\dot{X} + \frac{M_g}{M}\left(\frac{1}{X} - \frac{1}{L-X}\right)\dot{X}^2 \,, \quad (2)$$

which is by itself capable to determine the diathermal piston dynamics, once T(t) is expressed as a function of $\dot{x}(t)$ by means of the energy-conservation relation

$$\frac{1}{2}M\dot{x}^2 + 2nc_V T = 2nc_v T_0 \,. \quad (3)$$

If we assume the third term on the RHS to be much smaller than the first one, and, analogously, the first term on the LHS of Eq.(3) to be negligible with respect to the second one, Eq.(2) reduces to

$$\ddot{X} = \frac{nRT_0}{M}\left(\frac{1}{X} - \frac{1}{L-X}\right) - \sqrt{\frac{8nRM_g T_0}{\pi M^2}}\left(\frac{1}{X} + \frac{1}{L-X}\right)\dot{X} \,. \quad (4)$$

According to Eq.(4), the piston eventually attains the asymptotic equilibrium position X=L/2, corresponding to a vanishing velocity.

**Adiabatic piston**

The coupled system of equations predicts a final mechanical equilibrium position $X_f$ and temperatures $T_{Af}$ and $T_{Bf}$ dependent by the initial conditions X(0), $T_A(0)$ and $T_B(0)$, resulting in equal pressures in the two sections, [2],[7], i.e.,

$$\frac{T_{Af}}{X_f} = \frac{T_{Bf}}{L - X_f} \quad , \quad (5)$$

with $T_{Af} + T_{Bf} = T_A(0) + T_B(0) = 2T_0$ (energy conservation).

## 2a. Stochastic stage of diathermal-piston evolution

We wish now to investigate the stochastic motion of the diathermal piston occurring around its equilibrium position $X=L/2$ by adopting Langevin's approach. To this end, we introduce the displacement $x=X-L/2$ of the piston from the equilibrium position and assume, as verified *a posteriori*, that $|x| \ll L/2$. This allows us to expand the first term on the RHS of Eq.(4) around $X=L/2$ and set $X=L/2$ in the second term, thus obtaining, after adding the stochastic acceleration $a(t)$, the linear Langevin equation

$$\ddot{x} + \beta \dot{x} + \tilde{\omega}^2 x = a(t) \quad , \quad (6)$$

where

$$\beta = \frac{8N}{\sqrt{\pi}} \frac{v_0}{L} \varepsilon \quad , \quad (7)$$

$$\tilde{\omega}^2 = 4N \frac{v_0^2}{L^2} \varepsilon \quad , \quad (8)$$

and we have introduced the smallness parameter $\varepsilon = m/M$ and the most probable velocity $v_0 = (2kT_0/m)^{1/2}$ of the gas molecules at temperature $T_0$. The amplitude of the white-noise process $a(t)$ can be readily obtained by a straightforward application of the fluctuation-dissipation theorem (see, e.g., Ref. [6], pg.238), thus getting

$$<a(t)a(t')> = \frac{16N}{\sqrt{\pi}L} v_0^3 \varepsilon^2 \delta(t-t') \quad . \quad (9)$$

Equation (6) is formally identical to that describing the one dimensional Brownian motion of a harmonically-bound particle of mass M. [8] In the following, we will be dealing with the extreme "overdamped" regime $\beta \gg 2\tilde{\omega}$, corresponding to $\sqrt{M_g/M} \gg 1$ (see Eqs. (7), (8)).

In this case, the relevant average quantities $<x^2(t)>$ and $<\dot{x}^2(t)>$, where $<\ldots>$ stands for average over an ensemble of identical pistons starting from the common initial position $x=0$ and initial velocity $\dot{x} = 0$, read [8]

$$<x(t)^2> = \frac{kT_0}{M\tilde{\omega}^2}\{1 - \exp(-\beta t)[2\frac{\beta^2}{\tilde{\beta}_1^2}\sinh^2(\tilde{\beta}_1 t/2) + \frac{\beta}{\tilde{\beta}_1}\sinh(\tilde{\beta}_1 t) + 1]\}, \quad (10)$$

$$<\dot{x}(t)^2> = \frac{kT_0}{M}\{1 - \exp(-\beta t)[2\frac{\beta^2}{\tilde{\beta}_1^2}\sinh^2(\tilde{\beta}_1 t/2) - \frac{\beta}{\tilde{\beta}_1}\sinh(\tilde{\beta}_1 t) + 1]\}, \quad (11)$$

where $\tilde{\beta}_1 = \sqrt{\beta^2 - 4\tilde{\omega}^2} \cong \beta - 2\tilde{\omega}^2/\beta$.

Inspecting Eqs. (10) and 11) clearly reveals the existence of two characteristic times

$$\tilde{t}_{as} = \beta/2\tilde{\omega}^2 = \frac{1}{\sqrt{\pi}} \frac{L}{v_0}, \quad (12)$$

$$\tilde{t}_{th} = (1/2\beta) = \pi \frac{M}{M_g} \tilde{t}_{as} \ll \tilde{t}_{as}, \quad (13)$$

where $\tilde{t}_{th}$ represents a characteristic scale of piston velocity, corresponding to the time over which $<\dot{x}^2>$ reaches its asymptotic thermal value

$$<\dot{x}^2>_{as} = \frac{kT_0}{M}, \quad (14)$$

and $\tilde{t}_{as}$ is a characteristic scale of amplitude fluctuations, corresponding to the time over which $<x^2>$ reaches its asymptotic value

$$<x^2>_{as}^{(dp)} = \frac{kT_0}{M\tilde{\omega}^2} = \frac{1}{2N}\left(\frac{L}{2}\right)^2. \quad (15)$$

We note that Eq.(15) is consistent with the assumption upon which the linearized Eq.(6) is based since $x^2 \ll (L/2)^2$, and reproduces the standard result obtained in the frame of equilibrium-thermodynamic microscopic fluctuations.[9] Below (see Sects. 3 and 4), we will derive the result expressed by Eq. (15) both by a direct phase-space analysis and by maximizing Shannon's entropy for an ensemble of diathermal pistons.

**2b. Stochastic stage of adiabatic-piston evolution**

In the case of the adiabatic piston, the piston is moving along the equal-pressure line $T_A/X=T_B/(L-X)$ (see Eq.(5)) and, as a consequence, and unlike the diathermal case, the first term of Eq.(1) vanishes and the third one is no more negligible. Recalling Eq.(2), the stochastic equation of motion reads

$$\ddot{X} = -\sqrt{\frac{8nRM_gT}{\pi M^2}}(\frac{1}{X}+\frac{1}{L-X})\dot{X} + \frac{M_g}{M}(\frac{1}{X}-\frac{1}{L-X})\dot{X}^2 + a(t) \quad (16)$$

which, once linearized around $X=L/2$ and substituting $\dot{X}^2$ with its average thermal value $kT_0/M=(v_0^2/2)\varepsilon$ (two assumptions to be justified *a posteriori*) yields the Langevin equation

$$\ddot{x} + \beta\dot{x} + \omega^2 x = a(t), \quad (17)$$

where

$$\omega^2 = 2\frac{M_g}{M}\frac{1}{(L/2)^2}<\dot{x}^2>_{as} = 2\frac{M_g}{M}\frac{1}{(L/2)^2}\frac{kT_0}{M} = \varepsilon\tilde{\omega}^2. \quad (18)$$

By using Eqs. (10) and (11) with the substitutions of $\tilde{\omega}$ with $\omega$ and of $\tilde{\beta}_1$ with $\beta_1 = \sqrt{\beta^2 - 4\omega^2} \cong \beta - 2\omega^2/\beta$, we obtain

$$<x^2>^{(ap)}_{as} = \frac{kT_0}{M\omega^2} = \frac{1}{2N}\frac{M}{m}\left(\frac{L}{2}\right)^2 = \frac{1}{2}\frac{M}{M_g}\left(\frac{L}{2}\right)^2 \quad , \quad (19)$$

$$t_{as} = \beta/2\omega^2 = \frac{1}{\sqrt{\pi}}\frac{L}{v_0}\frac{M}{m}, \quad (20)$$

$$t_{th} = (1/2\beta) = \frac{\pi}{16N}t_{as} << t_{as}. \quad (21)$$

Note that the asymptotic displacement of the adiabatic piston is much larger than that of the diathermal piston,

$$\frac{<x^2>^{(ap)}_{as}}{<x^2>^{(dp)}_{as}} = \frac{M}{m} \quad , \quad (22)$$

a direct consequence of the smaller restoration force appearing in Eq.(17). However, if one assumes $M/M_g<<1$, $<x^2>^{(ap)}_{as}$ is still much smaller than $L/2$, as required by the linearization hypothesis. The validity of the hypothesis of fast thermalization of the piston velocity is supported by Eq.(21). Equation (22) implies values $\Delta V/V$ of the volume fluctuations of each sections of the cylinder (and thus of the corresponding $\Delta T/T$) much larger than those pertaining to equilibrium-thermodynamic microscopic fluctuations, so that the system cannot be considered in thermal equilibrium. On the other hand, the system is in a stationary regime, i.e., its probability distribution function does not depend on time. The dynamic evolution leading to this stationary regime corresponds to a Brownian motor-like stage occurring when the pressures in the two sections A and B become equal, i.e., the system in a state of marginal equilibrium along the equal-pressure line (see also 10,11]).

**3. Phase-space analysis of the diathermal piston**

Let us consider an ensemble of diathermal-piston systems, each consisting of an insulated cylinder, N molecules of mass m per section and a conducting frictionless piston. We introduce the 2N dimensional $\Gamma$-phase space whose points correspond to the dynamical states of the 2N gas molecules and look for the size $\Gamma(x)$ of the $\Gamma$-volume corresponding to a piston position between x and x+dx. To this aim, we observe that the constraint imposed by the piston position x on each of the molecules contained in the left (right) section implies the molecules to be confined between -L/2 and x (x and L/2), so that

$$\Gamma(x) = \left(\frac{L}{2} + x\right)^N \left(\frac{L}{2} - x\right)^N = \left(\frac{L}{2}\right)^{2N}\left(1 - 4\frac{x^2}{L^2}\right)^N \quad . \quad (23)$$

At this point, it is important to note the following features of the system. First, although the average distance among the molecules depends on the piston position x, the assumed perfect nature of the gas prevents any influence of x on the statistical distribution of molecular momentum coordinates.

Second, while the "particles" (basically, electrons and phonons), composing the diathermal piston, undergo mutual energy exchange, the average interaction of each of them with the totality of gas molecules is clearly negligible (compared to kT). Therefore, the dynamics of the total system (gas+piston) in phase-space is substantially described in the phase-space $\Omega$ associated with piston position and relative constraint on the gas molecules. In other words, the relevant phase-space volume $d\Omega$ corresponding to a piston position between x and x+dx is given by

$$d\Omega = (\frac{L}{2}+x)^N (\frac{L}{2}-x)^N dx = (\frac{L}{2})^{2N}(1-4\frac{x^2}{L^2})^N dx \quad , \quad (24)$$

i.e., in the limit $x^2/L^2 \ll 1$ and dropping the inessential constant $(L/2)^{2N}$,

$$d\Omega = \exp(-4Nx^2/L^2). \quad (25)$$

Since any member of our ensemble of systems is equally likely to be in any one of the various possible microstates, the time-independent normalized probability density p(x) of a piston being between x and x+dx is proportional to the corresponding phase-space volume, that is

$$p(x) = \sqrt{4N/\pi L^2}\exp(-4Nx^2/L^2), \quad (26)$$

which is in particular consistent with Eq.(15).
What happens for an ensemble of adiabatic pistons? In this case, as already noted, a "heavy particle" (the piston) simultaneously interacts with a very large number of gas molecules, so that its resulting average interaction energy cannot be considered negligible compared to the thermal energy kT. Besides, the interaction energy depends on the gas densities in section A and B, and thus on the position x. As a consequence, the argument leading to Eq.(24) does not apply and Eq.(26) does not hold. Therefore, it is not surprising that the random wandering of the adiabatic piston may result in an average displacement much larger than that of a diathermal piston. Conversely, if the number of particles is small enough to justify neglecting the piston interaction energy with respect to kT, the adiabatic- piston system (gas molecules plus piston) can be correctly regarded as a perfect gas, so that Eq.(26) is valid. This is actually the case for various molecular dynamics simulations. In particular, the results of Ref.[12], where the number of particles in each section is quite small (N=62) are in rather good agreement with Eq. (15), i.e., with Eq. (26). Similar numerical experiments involving a large number of particles, as would be required to test Eq.(19), do not appear to be presently available.

**4. Shannon's entropy for an ensemble of diathermal pistons**

We wish here to consider an ensemble of identical diathermal-piston systems and to find the distribution of maximal disorder. To this aim, we introduce the disorder D,

$$D = -\int P \ln P \, d\Omega \quad , \quad (27)$$

where $Pd\Omega$ is the frequency of systems whose representative position in the phase-space $\Omega$ is within $d\Omega$. In the present case (diathermal piston), we recall (see Sect.3) that the relevant (2N+1)-dimension phase-space $\Omega$ is defined through the relations

$$\Omega \to x, \{x_i\}, \{x_j\}, \ 1 \le i \le N, \ N+1 \le j \le 2N,$$
$$-L/2 \le x \le L/2, \ -L/2 \le x_i \le x, \ x \le x_j \le L/2, \quad (28)$$

where $x_i$ and $x_j$ respectively label the left and right molecule positions measured from X=L/2 (see Fig.1). The probability density P can be conveniently expressed as

$$P(x, \{x_i\}, \{x_j\}) = f(x) F_x(\{x_i\}, \{x_j\}) \ , \quad (29)$$

where f(x) is the piston position probability density and $F_x(\{x_i\}, \{x_j\})$ the 2N molecule-position *conditional* probability density. From Eqs. (27) and (29) we readily obtain

$$D = -\int_{-L/2}^{L/2} dx f(x) \ln f(x) + \int_{-L/2}^{L/2} dx f(x) \left[ -\int_{-L/2}^{x} \Pi dx_i \int_{x}^{L/2} \Pi dx_j F_x(\{x_i\}, \{x_j\}) \ln F_x(\{x_i\}, \{x_j\}) \right]$$
$$\equiv D_p + \int_{-L/2}^{L/2} dx f(x) D_g(x) \ . \quad (30)$$

Therefore, the total disorder D turns out to be the sum of the piston disorder $D_p$ and of the gas disorder $D_g(x)$ averaged over the piston position.

Let us know consider the standard Shannon entropy S of our ensemble, that is the *constrained maximum* of D.[9] More precisely, we wish to determine the distributions f(x) and $F_x(\{x_i\}, \{x_j\})$ that maximize D. To this end, it is convenient to proceed in two steps: first we look for the $F_x(\{x_i\}, \{x_j\})$ which maximizes $D_g(x)$ (for any fixed x), and then for the f(x) which maximizes D. For any fixed x, the distribution $F_x(\{x_i\}, \{x_j\})$ maximizing $D_g(x)$ is clearly uniform, i.e., independent from the $x_i$'s and the $x_j$'s, and yields for $D_g(x)$ the particular value $S_g(x)$ corresponding to the standard gas entropy, that is, apart from an inessential constant,

$$S_g(x) = \ln \left[ \frac{(L/2)^2 - x^2}{(L/2)^2} \right]^N = N \ln(1 - 4x^2/L^2) \ . \quad (31)$$

Thus, the system-ensemble entropy S corresponds to the maximum of

$$D = -\int_{-L/2}^{L/2} dx \, f(x) \ln f(x) + \int_{-L/2}^{L/2} dx \, f(x) S_g(x) \ , \quad (32)$$

over the normalized distributions f(x). In order to find the particular f(x) giving the above constrained maximum, we impose the vanishing of the variation

$$-\int_{-L/2}^{L/2} dx [1 + \ln f(x)] \partial f(x) + \int_{-L/2}^{L/2} dx \partial f(x) S_g(x) + \lambda \int_{-L/2}^{L/2} dx \partial f(x) \ , \quad (33)$$

where $\lambda$ is a Lagrange multiplier. In this way we obtain

$$1 + \ln f(x) = S_g(x) + \lambda \ , \quad (34)$$

equivalent to

$$f(x) = C \exp[S_g(x)] \ , \quad (35)$$

where C is a normalization constant. With the help of Eq.(31), Eq.(35) yields

$$f(x) = C (1 - 4x^2/L^2)^N \ , \quad (36)$$

i.e., for $x^2/L^2 \ll 1$,

$$f(x) = \sqrt{\frac{4N}{\pi L^2}} \exp(-4Nx^2/L^2) \ , \quad (37)$$

which coincides with Eq. (26).

## 5. Entropy variation

According to the results of Sect.2b, the adiabatic piston exhibits a rather intriguing behaviour associated with the presence of an asymptotic stationary regime characterized by fluctuations much larger than those corresponding to ordinary thermal equilibrium (e.g., the ones present in an otherwise identical diathermal piston). It can be assimilated to a Brownian particle in one dimension, whose associated state variable x(t) obeys the standard equation in the presence of an "internal" harmonic potential (see Eq.(17)). When considering the whole system gas-filled cylinder + adiabatic piston one can ask what are the consequences of the piston behaviour on the entropy variations $\Delta S$ of the system when starting from an initial condition of both thermal and mechanical equilibrium (piston held by latches at X=L/2, equal temperatures in the two sections). Since the adiabatic-piston ensemble eventually attains a stationary regime (time-independent probability distribution function), but the magnitude of the piston fluctuations does not allow to regard it as a thermodynamic-equilibrium regime, it is not trivial how to define $\Delta S$. A possible way out is to consider a system slightly different from the one dealt with in the previous Sections, that is one in which we insert a trap stopping the piston whenever its displacement equals the average asymptotic value $d \equiv [\langle x^2 \rangle_{as}^{(ap)}]^{1/2}$ given by Eq.(15): in this way, the final state is also of both thermal and mechanical equilibrium and the entropy variation of the system can be evaluated with the standard Clausius approach. More precisely, the entropy variation $\Delta S$ of the gas from the initial configuration to the final one is given by the change of the gas entropy between two states of common pressure and different volumes, i.e.,

$$\Delta S = \Delta S(A) + \Delta S(B) = nc_p [\ln \frac{(L/2)-d}{L/2} + \ln \frac{(L/2)+d}{L/2}] = nc_p \ln[1 - (\frac{d}{L/2})^2] \cong -k \frac{c_p}{2R} \frac{M}{m} \ , \quad (38)$$

where $c_p$ is the molar heat at constant pressure and advantage has been taken of Eq.(19).
The *negative* entropy variation expressed by Eq.(38) can be quite large, compared to the standard thermal equilibrium fluctuations $|\Delta S|_{diath}=kc_v/2R$, due to the presence of the factor M/m which is responsible for the large amplitude fluctuations of the piston position. Actually, the crucial open question is whether the constraint which eventually stops the piston at the position d can be considered a non-sentient Maxwell's demon.[13] In this respect, it is worth noting that our system differs from Feynman ratchet and, more in general, from Brownian motors, where the main mechanism is the rectification of *small* thermodynamic-equilibrium fluctuations.[14] We wish finally to stress that, as far as the adiabatic piston is concerned, the validity of our results is limited to *mesoscopic* regimes. More precisely, spatial dimensions and gas densities must be such that the time scales characterizing the relevant fluctuations are not so large to make unrealistic the whole process (see Eqs. (20) and (21)). For example, if one assumes the two cavities to be cubic and filled with a perfect gas in standard conditions, this requirement results in a length L of about one micron.[5]

**6. Conclusions**

We have revisited the stationary-regime reaching dynamics of the adiabatic-piston system and compared it with that of an otherwise identical diathermal-piston system, by using in both cases the standard Langevin approach. The second case is also dealt with in the frame of a straightforward phase-space analysis and reproduces the same result as Langevin's method, a fact that corroborates the validity of the adoption of this approach also when dealing with the adiabatic piston. Whether or not the relatively large fluctuations of the adiabatic piston (as compared to the diathermal one) could be harnessed is an open question.

# Figure 1

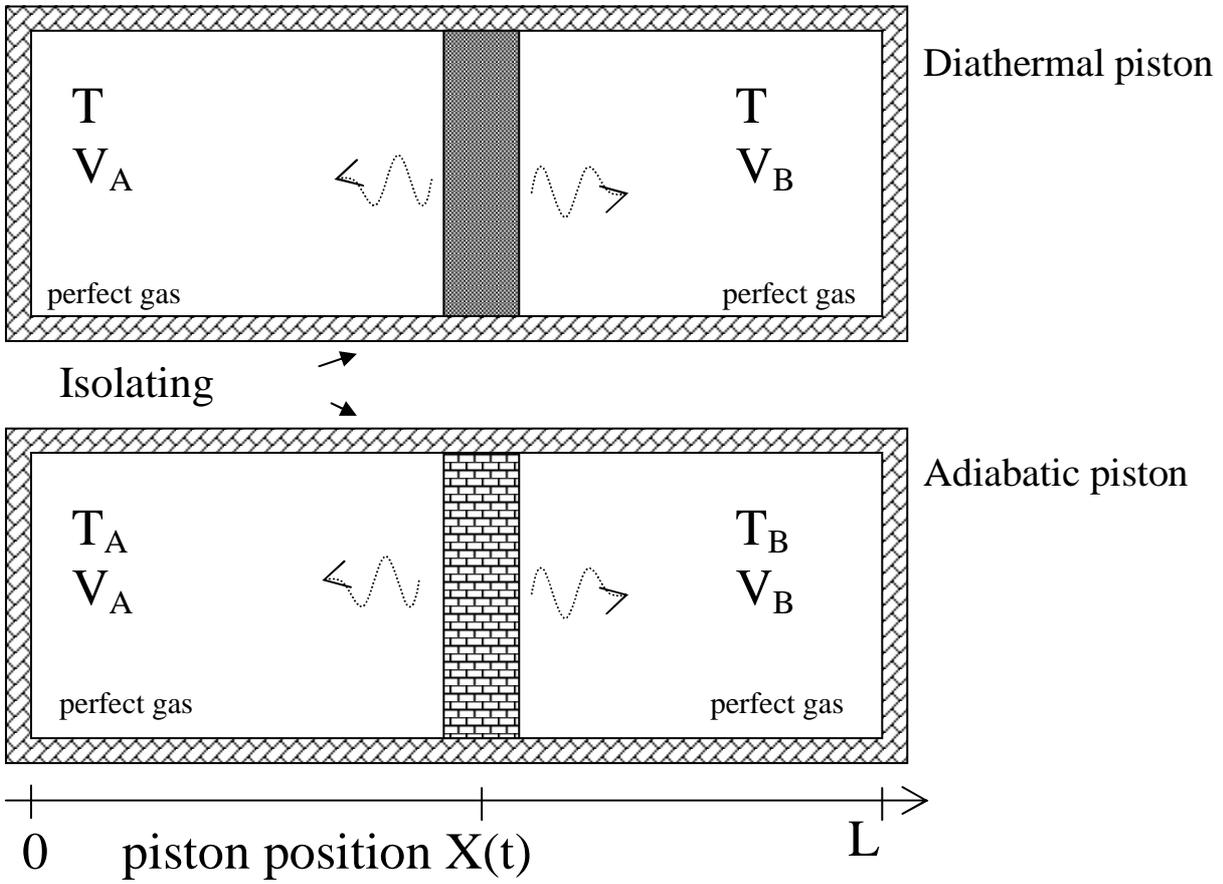